\documentclass[aps,prb,twocolumn,epsf,floatfix]{revtex4}
\usepackage{epsfig}
\usepackage{color}
\usepackage{graphicx}
\begin{document}


\title{Fano-Kondo effect in a two-level system with triple quantum dots:
shot noise characteristics}

\author{Tetsufumi~Tanamoto, Yoshifumi Nishi, and Shinobu Fujita}

\affiliation{Advanced LSI laboratory, Corporate R\&D Center,
Toshiba Corporation,\\ 1, Komukai Toshiba-cho, Saiwai-ku,
Kawasaki 212-8582, Japan}

\begin{abstract}
We theoretically compare transport properties of Fano-Kondo effect 
with those of Fano effect.  
We focus on shot noise characteristics of a triple quantum dot (QD) system
in the Fano-Kondo region at zero temperature, and discuss the effect of 
strong electric correlation in QDs. 
We found that the modulation of the Fano dip is strongly affected by the 
on-site Coulomb interaction in QDs. 
\end{abstract}
\maketitle

Quantum dot (QD) systems have attracted a lot of interest 
over many years because of their variety of controllability 
of small number of electrons in order to understand many-body 
effects in electronic systems.
Quantum correlation between localized states 
in QDs and free electrons in electrodes induces 
interesting phenomena such as the Fano effect and the Kondo effect.
A number of important experiments have been carried 
out\cite{Fano,Tarucha,Gores,Sato,Kobayashi,Otsuka,Rushforth,Sasaki}
and many theories
have been proposed\cite{Kang,Aligia,Wu,triple,TanaNishi}.
The Fano effect occurs as a result of quantum interference between 
a discrete energy state and a continuum state\cite{Fano}.
The Kondo effect is observed as a result of many-body correlations 
where internal spin degrees of freedom play 
an important role\cite{Tarucha}. The Fano-Kondo effect, which is a combination 
of the Fano effect and the Kondo effect, can be 
observed when on-site Coulomb interaction in a 
QD is strong\cite{Sato}. A T-shaped QD is considered to be 
suitable for discussing the Fano-Kondo 
effect\cite{Sato,Kobayashi,Otsuka,Kang,Aligia,Wu}. 
 
Quantum and thermal fluctuations are main obstacles
for observing quantum correlations, and are estimated 
through current noise characteristics.
Shot noise is a zero frequency limit of noise power spectrum
and provides various information on correlation 
of electrons. For uncorrelated electrons, shot noise $S_I$ 
shows Schottky result $S_I=2eI$ where $e$ is an electronic 
charge and $I$ is an electric current. The ratio of 
shot noise $S_I$ and full Poisson noise $2eI$
($I$ is an average current), $\gamma \equiv S_I/(2eI)$, 
is called the Fano factor, 
and indicates important noise properties. 
 
We have theoretically investigated transport properties of the triple 
QD system depicted in Fig.1,  
where QDs $a$ and $b$ are connected 
to electrodes through QD $d$\cite{TanaNishi}. 
This triple-QD system is considered to be in the same category as
the T-shaped QD. 
When coupling between QD $a$ and $b$ is larger than 
that between QD $b$ and $d$ ($t_C > t_d$), 
we can use this setup as apparatus for detecting  
two-level system (QD $a$ and QD $b$)
by a QD $d$ with electrodes
(Hereafter we call QD $d$ a detector QD). 
Moreover, when the number of electrons is controlled, 
double QD $a$ and $b$ can be regarded as a charge qubit\cite{tana0,Gilad}
with a Fano interference detector QD.
In Ref.\cite{TanaNishi}, we have shown that 
the Fano dip is modulated for a slow detector 
with no on-site Coulomb interaction in QD $d$.
However, noise properties, which are considered to be 
related to decoherence, has not been clarified.
Although $1/f$ noise induced by undesired trap sites is shown  
to be the largest cause of decoherence\cite{Astafiev},
shot noise is also a measure of decoherence in solid-state systems.

Wu {\it et al}.\cite{Wu} calculated noise properties of T-shaped QD system and showed
that shot noise strongly depends on the coupling strength between 
a side QD and a detector QD.
As tunneling coupling between side QD and detector QD increases, 
$\gamma$ quickly increase up to the Poisson value ($\gamma=1$).
L\'{o}pez {\it et al.} calculated shot noise of serially and laterally 
coupled double QD system and showed that $\gamma$ 
strongly depends on the coupling strength between QDs\cite{Lopez}.
Thus, $\gamma$ and shot noise reflect the 
coupling configuration of QD system and 
provide important information about the electronic 
structure of the system.
\begin{figure}
\includegraphics[width=6cm]{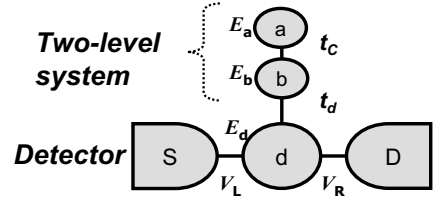}
\caption{Schematic plot of triple QD system. 
QDs $a$ and $b$ constitute a two-level system that is  
coupled to QD $d$ only which is connected to 
the electrodes. 
}
\label{System}
\end{figure}

Here, we compare zero temperature shot noise properties 
of the Fano-Kondo effect with those of the Fano effect, 
in order to reveal the effect of strong on-site Coulomb 
interaction on the transport properties.
The former case has stronger constraint 
than the latter case.
We assume an infinite Coulomb interaction 
for QD $a$ and $b$ and no Coulomb interaction for QD $d$
($U_a=U_b=\infty$, $U_d=0$) for the Fano-Kondo case. 
For the Fano case, 
we consider that there is no on-site Coulomb interaction for all QDs 
($U_a=U_b=U_d=0$). This corresponds to 
a case in which there is one degree of freedom 
\cite{Otsuka,Mahan} such that QDs are large 
without a spin scattering. 
For simplicity, we assume that there is a single energy level in each QD and 
that the two energy levels of  QD $a$ and QD $b$ coincide 
and correspond to gate voltages that are applied to those QDs.
We use slave-boson mean-field theory (SBMFT) based on the nonequilibrium Keldysh 
Green's function method.  The formulation 
of SBMFT is very useful and a good starting point 
for studying the transport properties 
of a strongly correlated QD system, 
although this method is usable at a lower temperature 
($T$) region than the Kondo temperature $T_K$\cite{Newns,Lopez}.

{\it Formulation---.}
Hamiltonian is constructed from electrode parts, QD parts, tunneling parts 
between an electrode and a QD, and those between QDs. 
For the Fano-Kondo case, additional constraint is required. 
The mean-field Hamiltonian for the Fano-Kondo case is described in terms 
of slave-bosons $b_{\alpha_1}$ 
$(\alpha_1=a,b)$ as:  
\begin{widetext}
\begin{eqnarray}
\lefteqn{
H^{\rm FK}=\sum_{\alpha=L,R}\sum_{k_\alpha,s}
E_{k_\alpha}c_{k_\alpha s}^\dagger c_{k_\alpha s}
+\sum_{\alpha_1=a,b,d}\sum_{s}
E_{\alpha_1}f_{\alpha_1s}^\dagger f_{\alpha_1s}
+\sum_{\alpha_1=a,b}
\lambda_{\alpha_1} \left(\sum_s f_{\alpha_1s}^\dagger f_{\alpha_1s}
+b_{\alpha_1}^\dagger b_{\alpha_1}-1\right)
}\nonumber \\
\!&+&\!\frac{t_C}{N}\sum_{s}(
 f_{as}^\dagger b_ab_b^\dagger f_{bs}
+f_{bs}^\dagger b_bb_a^\dagger f_{as})
+\frac{t_d}{N}\sum_{s}(
 f_{ds}^\dagger b_b^\dagger f_{bs}
+f_{bs}^\dagger b_b f_{ds})
+\sum_{\alpha=L, R}\frac{V_\alpha}{\sqrt{N}} \sum_{k_\alpha,s}
(c_{k_\alpha s}^\dagger f_{ds}
+f_{ds}^\dagger c_{k_\alpha s})
\label{Hamiltonian}
\end{eqnarray}
\end{widetext}
where $E_{k_\alpha}$ is the energy level for source ($\alpha=L$) and drain ($\alpha=R$) 
electrodes. $E_{a}$, $E_{b}$ and $E_{d}$ are energy levels for the three QDs, respectively.
$t_C$, $t_d$ and $V_\alpha$ are the tunneling coupling strength 
between QD $a$ and QD $b$, that between QD $b$ and QD $d$, and 
that between QD $d$ and electrodes, respectively.
$c_{k_\alpha s}$ and $f_{\alpha_1s}$ are annihilation operators of the electrodes, 
and of the three QDs $(\alpha_1=a,b,d)$, respectively.
$s$ is spin degree of freedom with spin degeneracy $N$; here we apply $N=2$.
$\lambda_{\alpha_1}$ is a Lagrange multiplier.
We take $z_{\alpha_1} \equiv b_{\alpha_1}^\dagger b_{\alpha_1}/2$ and 
$\tilde{E}_{\alpha_1} \equiv E_{\alpha_1}+\lambda_{\alpha_1}$
as mean-field parameters for QD $a$ and QD $b$. 
The Hamiltonian for the Fano case is similar to $H^{\rm FK}$ except 
that $\lambda_a=\lambda_b=0$ and $b_a=b_b=1$ in Eq.(\ref{Hamiltonian}).

In the Fano-Kondo effect, four self-consistent equations to determine mean-field 
parameters $\lambda_{\alpha_1}$ and $b_{\alpha_1}$ ($\alpha_1=a,b$) are 
required and expressed as
\begin{eqnarray}
& &\tilde{t}_C \sum_{s}
\langle f_{bs}^\dagger  f_{as} \rangle 
\!+\!\lambda_a |b_a|^2 =0, 
\label {s_eq1_1}\\
& &\tilde{t}_C \sum_{s}
\langle f_{as}^\dagger f_{bs} \rangle
+\tilde{t}_d \sum_{s}
\langle f_{ds}^\dagger f_{bs} \rangle
\!+\!\lambda_b |b_b|^2 =0, 
\label {s_eq1_2}\\
& & \sum_s \langle f_{\alpha_1 s}^\dagger f_{\alpha_1 s}\rangle 
+|b_{\alpha_1 }|^2=1, 
\ \ (\alpha_1=a,b),
\label {s_eq1_3}
\end{eqnarray}

Current and noise formula are expressed by Keldysh Green's functions. 
Keldysh Green's functions are obtained by applying analytic continuation rules to 
the equations of motion, which are derived from  
the above Hamiltonian\cite{Newns,Lopez}. 
For example, Green's functions for QDs are given as  
$
G_{aa}^r(\omega)=
[(\omega-\tilde{E}_b)B_r -|\tilde{t}_d|^2]/B_{00}
$, 
$
G_{bb}^r(\omega)=
[(\omega-\tilde{E}_a)B_r/B_{00}
$ and 
$
G_{dd}^r(\omega) =D_{ab}/B_{00}
$
{\it etc.}, 
where 
$D_{ab}\equiv (\omega-\tilde{E}_a)(\omega-\tilde{E}_b)
-\tilde{t}_c^2$, 
$B_r\equiv \omega-\tilde{E}_d+i\Gamma$ and 
$B_{00} \equiv {D_{ab}B_r-(\omega-\tilde{E}_a)|\tilde{t}_d|^2}$
with 
$\tilde{t}_C=t_C b_a b_b^\dagger/N$ and
$\tilde{t}_d=t_d b_b^\dagger/N$.
Here, $\Gamma_\alpha\equiv 2\pi \rho_\alpha (\mu_\alpha)|V_\alpha |^2$ is the tunneling rate 
between $\alpha$ electrode and QD $d$ with a density of states (DOS), 
$\rho_\alpha (\mu_\alpha)$, for each electrode at Fermi energy $\mu_\alpha$. 
$\Gamma\equiv(\Gamma_L+\Gamma_R)/2$ and 
we assume $\Gamma_L=\Gamma_R$.

Source current $I_L$ is expressed as
\begin{equation}
I_L=\frac{2e}{h} \int_{-\infty}^{\infty} d\omega T(\omega)
(f_L(\omega)-f_R(\omega) )
\end{equation}
where transmission probability $T(\omega)$ is given as
\begin{equation}
T(\omega)=\frac{\Gamma_L\Gamma_R |D_{ab}|^2}{
[D_{ab} (\omega-E_d)-(\omega-\tilde{E}_a)z_a t_d^2/2]^2
+\Gamma^2 D_{ab}^2/4
}  
\label{trans}
\end{equation}
(the denominator is $B_{00}$)
and Fermi distribution functions 
$f_\alpha (\omega)\equiv [\exp ((\omega-\mu_\alpha)/T)+1]^{-1}$ where 
we set symmetrical bias condition: 
$\mu_L=E_F-eV/2$ and $\mu_R=E_F+eV/2$ with $E_F=0$.
Note that in the present case we can check that 
$I_L$ and $I_R$ are symmetric and satisfy a 
current conservation.
Conductance is given as $G=-\frac{2e}{h}\int d\omega
T(\omega) \frac{\partial f_L(\omega) }{\partial\omega}
$.
The transmission probability is related to a DOS 
of the detector QD $\rho_d (\omega)=-{\rm Im}G_{dd}^r (\omega)/\pi$ such as
$
T(\omega)=\frac{2\Gamma_L\Gamma_R}{\Gamma_L+\Gamma_R} \pi \rho_d(\omega),
$
which means that we can discuss characteristics of DOS similar to a transmission 
probability.

Current noise is calculated as a correlation function 
of current fluctuation as 
$
S(t,t')=\frac{1}{2}$ $[\langle \{ \hat{I}_L(t),\hat{I}_L(t') \} \rangle 
-2\langle \hat{I}_L(t) \rangle^2 ],
$
where $\hat{I}_L(t)=(ie/\hbar )\sum (V_L/\sqrt{N}) 
[c_{k_L s}^\dagger(t) f_{ds}(t)-\mbox{H.c.}]$ is a current operator. 
A derivation procedure similar to that in Ref.\cite{Lopez} is applied to our case,
we obtained noise formula at $T=0$ as
\begin{equation}
S(V)=\frac{4e^2}{h} \int_{-eV/2}^{eV/2} d\omega T(\omega) (1-T(\omega)).
\end{equation} 
The Fano factor $\gamma$ at zero bias $V=0$ is obtained by $\gamma=1-T(E_F)$, 
and indicates that shot noise is in the sub-Poissonian region ($\gamma<1$).
Similar to Ref.\cite{TanaNishi}, we classify our triple QD system by magnitude 
of $t_C/t_d$ and $\Gamma/t_d$. 
The ratio $t_C/t_d$ compares the internal coupling strength in a two-level system
with that between the two-level system and the detector, 
and we regard the case where $t_C/t_d=5$ as  
a strongly coupled two-level system and the case where $t_C/t_d=1$
as a weakly coupled two-level system.
If $\Gamma /t_d$ is large, the electron 
that flows through QD $d$ is so fast that it cannot detect the oscillation of 
an electron in the coupled QDs $a$ and $b$.
If $\Gamma /t_d$ is small, the electron 
that flows through QD $d$ can observe the 
evidence of bonding and antibonding 
states.
We call a detector with large $\Gamma/t_d=2$ a fast detector,
and one with smaller $\Gamma/t_d=0.4$ a slow detector.
We assume that $D= 20 t_d$, $ |E_d| <0.4t_d$, $\Gamma>0.4t_d$ 
and $E_F=0$ ($D$ is a bandwidth). Then, we have  
$T_K\sim D e^{-\pi |\tilde{E}_d-E_F|/\Gamma}$ 
$\sim 1.6 t_d$.

\begin{figure}
\includegraphics[width=7.5cm]{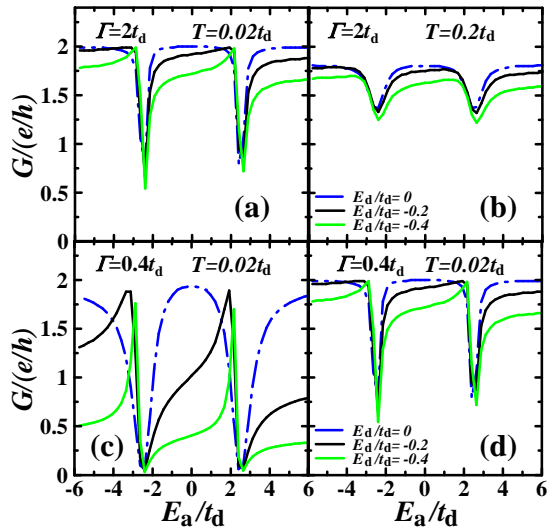}
\caption{Conductance $G$ as a function of an energy level
of the two-level system $E_a (=E_b)$ with 
a strong coupling ($t_C/t_d=5$) for the Fano case ($U_a=U_b=0$).
(a) $T/t_d=0.02$ and (b) $T/t_d=0.2$ for a fast detector 
($\Gamma/t_d=2$).
(c) $T/t_d=0.02$ and (d)$T/t_d=0.2$ for a slow detector 
($\Gamma/t_d=0.4$).
$E_F=0$. }
\label{Strong}
\end{figure}
\begin{figure}
\includegraphics[width=7cm]{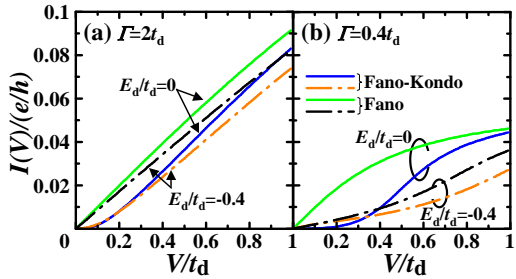}
\caption{$I$-$V$ characteristics for a strong coupling ($t_C/t_d=5$) at $T=0$.
(a) Fast detector ($\Gamma/t_d=2$) and (b) slow detector ($\Gamma/t_d=0.4$).
$E_a=E_b=0=E_F$.}
\label{Cond_strong}
\end{figure}

\begin{figure}
\includegraphics[width=7cm]{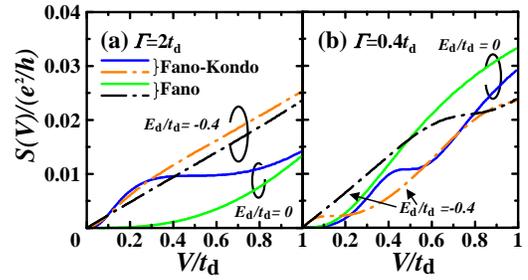}
\caption{Shot noise as a function of bias voltage 
for a strong coupling ($t_C/t_d=5$).
(a) Fast detector($\Gamma/t_d=2$). (b) Slow detector($\Gamma/t_d=0.4$). 
$T=0$. $E_a=E_b=0=E_F$. }
\label{SN_strong}
\end{figure}

\begin{figure}
\includegraphics[width=7.5cm]{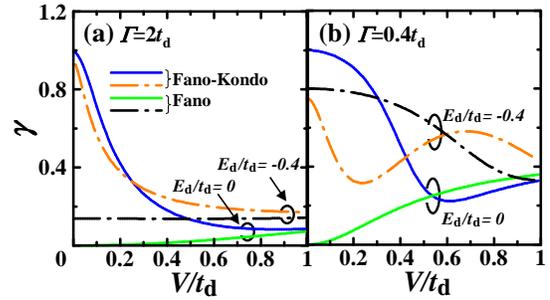}
\caption{Fano factor $\gamma$ as a function of bias voltage.
$E_a =E_b=0=E_F$ for a strong coupling ($t_C/t_d=5$).
(a) Fast detector($\Gamma/t_d=2$) and (b) slow detector($\Gamma/t_d=0.4$). }
\label{FF_strong}
\end{figure}

\begin{figure}
\includegraphics[width=7.5cm]{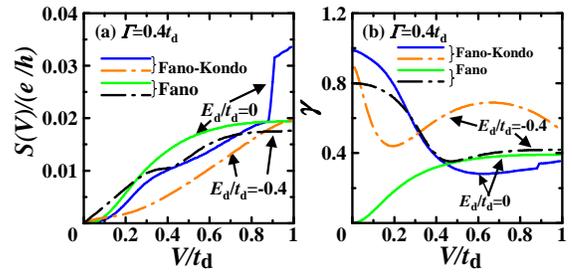}
\caption{Weak coupling case($t_C/t_d=1$) for a 
slow detector ($\Gamma/t_d=0.4$). 
(a) Shot noise and (b) Fano factor $\gamma$ as a function of bias voltage. 
$T=0$. $E_a=E_b=0=E_F$.}
\label{SN_weak}
\end{figure}

{\it Numerical calculations.}---
Here, we show numerical results of our triple QD system 
in the Fano-Kondo effect and the Fano effect.
First, Fig. \ref{Strong} shows conductance of the Fano case as a function of $E_a$.
We can see a clear double-peak structure in every figure. 
This is in a large contrast with 
our previous results of the Fano-Kondo effects(Ref.\cite{TanaNishi}) 
where modulation of a single Fano dip can be seen only by a slow detector
($\Gamma=0.4t_d$) at low temperature ($T=0.02t_d$).
The present clear double-peak structure is a direct result of 
the form of transmission probability $T(\omega)$, in particular,
$D_{ab}$ in the numerator of Eq.(\ref{trans}).
These results show that Kondo effect, spin exchange effect, 
greatly changes the Fano effect.
Dip structure is the largest for a slow detector at low temperature (Fig.\ref{Strong}(c)).
The asymmetry of the dip structure for $E_d \neq 0$ 
can also be understood from Eq.(\ref{trans}). 
Because $G \propto T(\omega \sim 0)$ at low temperature and we set $E_a=E_b$, 
we have $D_{ab}(\omega=0)=\tilde{E}_a^2-\tilde{t}_c^2$. Thus, for $E_d=0$, both 
the numerator and the denominator of $T(\omega)$ are symmetric for $E_a$. 
However, when $E_d\neq 0$,  the denominator deviates from symmetric form because of 
the $D_{ab} (\omega-E_d)$ in the expression. 

Figure \ref{Cond_strong} shows current-voltage 
($I$-$V$) characteristics at $T=0$.
We can see that all current looks similar for a fast detector ((a)) 
for both the Fano effect and the Fano-Kondo effect.
This indicates that a fast detector is less sensitive to 
quantum states of QD system than a slow detector.
Current of the Fano-Kondo effect is always less than that of 
the Fano effect. This indicates that stronger electronic correlation 
in QD $d$ suppresses its current.

Figure \ref{SN_strong} shows shot noise characteristics 
as a function of bias voltage across the detector QD.
$E_d$ dependence is simpler for a fast detector ((a)).
This is because a fast detector is 
more sensitive to energy level of a detector QD $d$ than 
energy levels in two-level system. 
Shot noise of a slow detector reflects internal states 
of two-level system.

Although magnitude of current for a fast detector is 
larger than that for a slow detector (Fig. \ref{Cond_strong}), 
magnitude of shot noise for a fast detector is 
of the same order as that of a slow detector (Fig. \ref{SN_strong}). 
Thus, $\gamma$ for a slow detector is relatively larger than that 
for a fast detector as shown in Fig. \ref{FF_strong}.
This is because of the stronger coupling of flowing electrons 
with two-level states in a slow detector.
Strong nonlinearity can be seen around $V=0$, because   
energy levels of QDs are close to a Fermi energy 
($E_a=E_b=0=E_F$) and strongly coupled with electrode electrons.
In both a fast detector and a slow detector, 
$\gamma$ for the Fano-Kondo case is larger than 
that for the Fano case. This indicates that 
stronger electronic correlation induces more noise.

Figures \ref{SN_weak}(a) and (b)  show shot noise and $\gamma$ 
of weak coupling ($t_C/t_d=1$) for a slow detector ($\Gamma/t_d=0.4$).
Similar to Fig. \ref{SN_strong} (b), shot noise is modulated 
by changing $E_d$ reflecting a two-level state.
Compared with Fig.\ref{SN_strong}, Fig. \ref{SN_weak}(a) 
shows that modulation by $E_d$ becomes more complicated. 
This is because energy levels of weakly coupled triple QDs are 
more close with each other than those of strongly coupled QDs. 
Figure \ref{SN_weak} (b) shows  $\gamma$ characteristics for a weakly coupled 
slow detector.
We can see that values of $\gamma$ become closer with each other, 
reflecting closer coupling between QDs.

{\it Discussion.}---
Our numerical results show that, as coupling in a two-level state (qubit) 
becomes stronger, noise increases. In addition, 
a slower detector, which is found to be more desirable for reading out 
a two-level state, induces more noise than a fast detector.
Thus,  there is a trade-off in that reading out more detailed information 
induces more noise or larger Fano factor. 
Appropriate parameters ($t_C/t_d$, $\Gamma/t_d$ etc.) should 
be determined depending on sensitivity of the external circuit 
connected to this triple QD system.
As noted in the introduction, in order to use the two-level 
state as a qubit, a stronger constraint is required so that one excess electron
stays in the two-level system. This would be realizable, for example, 
by forming smaller and closely coupled QDs, such that two electrons 
are not permitted into the QDs because of their repulsive Coulomb 
interaction. As shown in the numerical results, the Fano factor 
for stronger correlation (Fano-Kondo case) is larger 
than that for weaker correlation (Fano case). Thus, 
it is possible that we will have to accept larger back-action when 
introducing charge qubit condition.
More elaborate control of the measurement setup would be required 
for a charge qubit system. 

Kobayashi {\it et al.}\cite{Kobayashi} discussed the rapid smearing out of 
the dip structure with increasing temperature mainly owing to 
thermal broadening. 
Although we assume one energy level in each QD at $T=0$, if we take 
more energy levels in each QD into consideration, the Fano dip would 
smear out rapidly with increasing temperature.


In conclusion, 
focusing on the Fano effect and the Fano-Kondo effect in a two-level state, 
we studied noise properties of a triple QD system.
We have shown that, depending on the coupling strength among the triple QDs, 
noise and the Fano factor are greatly modulated for a slow detector.
In particular, we found that detailed reading of a two-level state 
is inclined to increase noise characteristics of the system.  

We are grateful to A. Nishiyama, J. Koga,   
R. Ohba, T. Otsuka and M. Eto for valuable discussions.

\end{document}